\documentclass[aps,prl,reprint,notitlepage,showpacs,nofootinbib,superscriptaddress,twocolumn
,nofloats,preprintnumbers]{revtex4-1}
\usepackage{graphicx}
\usepackage[utf8]{inputenc} 
\usepackage{amsmath}
\usepackage{amssymb}
\usepackage{ulem}
\usepackage{float}
\usepackage{comment}
\usepackage{slashed}

\usepackage[T1]{fontenc} 
\usepackage{parskip}
\usepackage{bm}         
\usepackage{array}
\usepackage{multirow}   
\usepackage{booktabs}   

\usepackage{textcomp}
\usepackage{amssymb}
\usepackage{ulem}
\usepackage{cancel}

\usepackage{enumerate}
\usepackage{caption}
\usepackage{subcaption}
\usepackage{ragged2e}
 \DeclareCaptionJustification{justified}{\justifying}
\usepackage{lineno}

\usepackage{amsmath}
\usepackage{amssymb}
\usepackage{indentfirst}
\usepackage[usenames,dvipsnames]{xcolor}
\usepackage[colorlinks=true,citecolor=darkred,urlcolor=darkred, pdfborder={0 0 0}]{hyperref}
\definecolor{darkred}{rgb}{0.6,0,0}

\definecolor{linkcolor}{rgb}{0,0,0.5}
 
\newcommand {\ignore}[1]{}

\definecolor{bostonuniversityred}{rgb}{0.8, 0.0, 0.0}

\def\gsim{\raise0.3ex\hbox{$\;>$\kern-0.75em\raise-1.1ex\hbox{$\sim\;$}}}
\def\lsim{\raise0.3ex\hbox{$\;<$\kern-0.75em\raise-1.1ex\hbox{$\sim\;$}}}

\usepackage{wrapfig}

\usepackage{soul}
\definecolor{mightnightblue}{RGB}{25,25,112}

\definecolor{brown}{rgb}{0.59, 0.29, 0.0}

\def\21{$\mathrm{SU(2)_L \otimes U(1)_Y}$}

\begin{document}

\title{Search for inelastic dark matter-nucleus scattering with the PICO-60 CF$_{3}$I and C$_{3}$F$_{8}$ bubble chambers}

\author{E.~Adams}
\affiliation{Department of Physics, Queen's University, Kingston, K7L 3N6, Canada}

\author{B.~Ali}
\affiliation{Institute of Experimental and Applied Physics, Czech Technical University in Prague, Prague, Cz-12800, Czech Republic}

\author{I.~J.~Arnquist}
\affiliation{Pacific Northwest National Laboratory, Richland, Washington 99354, USA}

\author{D.~Baxter}
\affiliation{Fermi National Accelerator Laboratory, Batavia, Illinois 60510, USA}

\author{E.~Behnke}
\affiliation{Department of Physics, Indiana University South Bend, South Bend, Indiana 46634, USA}

\author{M.~Bressler}
\affiliation{Department of Physics, Drexel University, Philadelphia, Pennsylvania 19104, USA}

\author{B.~Broerman}
\affiliation{Department of Physics, Queen's University, Kingston, K7L 3N6, Canada}

\author{C.~J.~Chen}
\affiliation{
Department of Physics and Astronomy, Northwestern University, Evanston, Illinois 60208, USA}

\author{K.~Clark}
\affiliation{Department of Physics, Queen's University, Kingston, K7L 3N6, Canada}

\author{J.~I.~Collar}
\affiliation{Enrico Fermi Institute, KICP, and Department of Physics,
University of Chicago, Chicago, Illinois 60637, USA}

\author{P.~S.~Cooper}
\affiliation{Fermi National Accelerator Laboratory, Batavia, Illinois 60510, USA}

\author{C.~Cripe}
\affiliation{Department of Physics, Indiana University South Bend, South Bend, Indiana 46634, USA}

\author{M.~Crisler}
\affiliation{Fermi National Accelerator Laboratory, Batavia, Illinois 60510, USA}

\author{C.~E.~Dahl}
\affiliation{
Department of Physics and Astronomy, Northwestern University, Evanston, Illinois 60208, USA}
\affiliation{Fermi National Accelerator Laboratory, Batavia, Illinois 60510, USA}

\author{M.~Das}
\affiliation{High Energy Nuclear \& Particle Physics Division, Saha Institute of Nuclear Physics, Kolkata, India}


\author{S.~Fallows}
\affiliation{Department of Physics, University of Alberta, Edmonton, T6G 2E1, Canada}

\author{J.~Farine}
\affiliation{School of Natural Sciences, Laurentian University, Sudbury, ON P3E 2C6, Canada}
\affiliation{SNOLAB, Lively, Ontario, P3Y 1N2, Canada}
\affiliation{Department of Physics, Carleton University, Ottawa, Ontario, K1S 5B6, Canada}

\author{R.~Filgas}
\affiliation{Institute of Experimental and Applied Physics, Czech Technical University in Prague, Prague, Cz-12800, Czech Republic}

\author{A. Garc\'{\i}a-Viltres}
\email[Corresponding: ]{agarciaviltres@gmail.com}
\affiliation{Instituto de F\'isica, Universidad Nacional Aut\'onoma de M\'exico, M\'exico D.\:F. 01000, M\'exico}

\author{G.~Giroux}
\affiliation{Department of Physics, Queen's University, Kingston, K7L 3N6, Canada}

\author{O.~Harris}
\affiliation{Northeastern Illinois University, Chicago, Illinois 60625, USA}

\author{T.~Hillier}
\affiliation{School of Natural Sciences, Laurentian University, Sudbury, ON P3E 2C6, Canada}

\author{E.~W.~Hoppe}
\affiliation{Pacific Northwest National Laboratory, Richland, Washington 99354, USA}

\author{C.~M.~Jackson}
\affiliation{Pacific Northwest National Laboratory, Richland, Washington 99354, USA}

\author{M.~Jin}
\affiliation{
Department of Physics and Astronomy, Northwestern University, Evanston, Illinois 60208, USA}

\author{C.~B.~Krauss}
\affiliation{Department of Physics, University of Alberta, Edmonton, T6G 2E1, Canada}

\author{V.~Kumar}
\affiliation{High Energy Nuclear \& Particle Physics Division, Saha Institute of Nuclear Physics, Kolkata, India}

\author{M.~Laurin}
\affiliation{D\'epartement de Physique, Universit\'e de Montr\'eal, Montr\'eal, H2V 0B3, Canada}

\author{I.~Lawson}
\affiliation{School of Natural Sciences, Laurentian University, Sudbury, ON P3E 2C6, Canada}
\affiliation{SNOLAB, Lively, Ontario, P3Y 1N2, Canada}

\author{A.~Leblanc}
\affiliation{School of Natural Sciences, Laurentian University, Sudbury, ON P3E 2C6, Canada}

\author{H.~Leng}
\affiliation{Materials Research Institute, Penn State, University Park, Pennsylvania 16802, USA}

\author{I.~Levine}
\affiliation{Department of Physics, Indiana University South Bend, South Bend, Indiana 46634, USA}

\author{C.~Licciardi}
\affiliation{School of Natural Sciences, Laurentian University, Sudbury, ON P3E 2C6, Canada}
\affiliation{SNOLAB, Lively, Ontario, P3Y 1N2, Canada}
\affiliation{Department of Physics, Carleton University, Ottawa, Ontario, K1S 5B6, Canada}

\author{W.~H.~Lippincott}
\affiliation{Fermi National Accelerator Laboratory, Batavia, Illinois 60510, USA}
\affiliation{Department of Physics, University of California Santa Barbara, Santa Barbara, California 93106, USA}



\author{P.~Mitra}
\affiliation{Department of Physics, University of Alberta, Edmonton, T6G 2E1, Canada}

\author{V.~Monette}
\affiliation{D\'epartement de Physique, Universit\'e de Montr\'eal, Montr\'eal, H2V 0B3, Canada}

\author{C.~Moore}
\affiliation{Department of Physics, Queen's University, Kingston, K7L 3N6, Canada}

\author{R.~Neilson}
\affiliation{Department of Physics, Drexel University, Philadelphia, Pennsylvania 19104, USA}

\author{A.~J.~Noble}
\affiliation{Department of Physics, Queen's University, Kingston, K7L 3N6, Canada}

\author{H.~Nozard}
\affiliation{D\'epartement de Physique, Universit\'e de Montr\'eal, Montr\'eal, H2V 0B3, Canada}

\author{S.~Pal}
\affiliation{Department of Physics, University of Alberta, Edmonton, T6G 2E1, Canada}

\author{M.-C.~Piro}
\affiliation{Department of Physics, University of Alberta, Edmonton, T6G 2E1, Canada}

\author{A.~Plante}
\affiliation{D\'epartement de Physique, Universit\'e de Montr\'eal, Montr\'eal, H2V 0B3, Canada}

\author{S.~Priya}
\affiliation{Materials Research Institute, Penn State, University Park, Pennsylvania 16802, USA}

\author{C.~Rethmeier}
\affiliation{Department of Physics, University of Alberta, Edmonton, T6G 2E1, Canada}

\author{A.~E.~Robinson}
\affiliation{D\'epartement de Physique, Universit\'e de Montr\'eal, Montr\'eal, H2V 0B3, Canada}

\author{J.~Savoie}
\affiliation{D\'epartement de Physique, Universit\'e de Montr\'eal, Montr\'eal, H2V 0B3, Canada}

\author{A.~Sonnenschein}
\affiliation{Fermi National Accelerator Laboratory, Batavia, Illinois 60510, USA}

\author{N.~Starinski}
\affiliation{D\'epartement de Physique, Universit\'e de Montr\'eal, Montr\'eal, H2V 0B3, Canada}

\author{I.~\v{S}tekl}
\affiliation{Institute of Experimental and Applied Physics, Czech Technical University in Prague, Prague, Cz-12800, Czech Republic}

\author{D.~Tiwari}
\affiliation{D\'epartement de Physique, Universit\'e de Montr\'eal, Montr\'eal, H2V 0B3, Canada}

\author{E.~V\'azquez-J\'auregui}
\email[Corresponding: ]{ericvj@fisica.unam.mx}
\affiliation{Instituto de F\'isica, Universidad Nacional Aut\'onoma de M\'exico, M\'exico D.\:F. 01000, M\'exico}

\author{U.~Wichoski}
\affiliation{School of Natural Sciences, Laurentian University, Sudbury, ON P3E 2C6, Canada}
\affiliation{SNOLAB, Lively, Ontario, P3Y 1N2, Canada}
\affiliation{Department of Physics, Carleton University, Ottawa, Ontario, K1S 5B6, Canada}

\author{V.~Zacek}
\affiliation{D\'epartement de Physique, Universit\'e de Montr\'eal, Montr\'eal, H2V 0B3, Canada}

\author{J.~Zhang}
\altaffiliation[now at ]{Argonne National Laboratory}
\affiliation{
Department of Physics and Astronomy, Northwestern University, Evanston, Illinois 60208, USA}

\collaboration{PICO Collaboration}
\noaffiliation

\date{\today}

\begin{abstract}

PICO bubble chambers have exceptional sensitivity to inelastic dark matter-nucleus interactions due to a combination of their extended nuclear recoil energy detection window from a few keV to $O$(100 keV) or more and the use of iodine as a heavy target. Inelastic dark matter-nucleus scattering is interesting for studying the properties of dark matter, where many theoretical scenarios have been developed. This study reports the results of a search for dark matter inelastic scattering with the PICO-60 bubble chambers. The analysis reported here comprises physics runs from PICO-60 bubble chambers using CF$_{3}$I and C$_{3}$F$_{8}$. The CF$_{3}$I run consisted of 36.8 kg of CF$_{3}$I reaching an exposure of 3415 kg-day operating at thermodynamic thresholds between 7 and 20 keV. The C$_{3}$F$_{8}$ runs consisted of 52 kg of C$_{3}$F$_{8}$ reaching exposures of 1404 kg-day and 1167 kg-day running at thermodynamic thresholds of 2.45 keV and 3.29 keV, respectively. The analysis disfavors various scenarios, in a wide region of parameter space, that provide a feasible explanation of the signal observed by DAMA, assuming an inelastic interaction, considering that the PICO CF$_{3}$I bubble chamber used iodine as the target material. 

\end{abstract}

\maketitle


\section{\label{sec:introduction}Introduction}

There is overwhelming evidence indicating that most of the matter in the Universe is non-baryonic \cite{EINASTO1974,Ostrike1974,Ostriker:1973uit,bd0fcfe34889413099a774ad5bded97a,10.2307/45312,Bahcall:2013epa,SDSS:2007umu,10.1093/mnras/stt572,Bennett_2013,Ade:1530672}. Searches for particle Dark Matter (DM) are underway with sensitive detectors in underground laboratories operating at $O$(keV) thresholds in ultra-low background environments \cite{PhysRevLett.127.261802,XENON:2018voc,HORN2015504,PhysRevLett.128.011801,Cooley_2010,Kluck_2020}. The typical expected signal is a nuclear recoil induced by the scattering of dark matter and the target nucleus. Detectors are currently sensitive to cross-sections as low as $10^{-47}$ cm$^2$ for scalar interactions (spin-independent) and $10^{-41}$ cm$^2$ for axial–vector (spin-dependent) interactions for masses between 10 and 100 GeV/$c^2$. In direct detection searches with spin-independent couplings, limits have been set assuming coherent elastic interactions between the dark matter particle and nuclei. Another interesting, viable and theoretically well-motivated  possibility, is an inelastic interaction. A possible scenario for this interaction includes a rich dark sector with multiple states, where the scattering induces a transition from a ground state into a heavier state. The simplest case would consist of only two states $\chi_1$ (lighter) and $\chi_2$ (heavier), where the mass splitting is $\delta=M_{\chi_2}-M_{\chi_1}$. Many well-motivated models have been proposed~\cite{Hall:1997ah,Arina:2007tm,An:2011uq,Barello:2014uda, Feng2022}. Inelastic dark matter has been suggested as a simple and elegant solution to the DAMA signal \cite{PhysRevD.79.043513,PhysRevD.64.043502,Schmidt_Hoberg_2009, PhysRevD.80.115008}. Phenomenological scenarios such as proton–philic spin-dependent \cite{Kang:2018zld}, inelastic scattering predominantly coupling to the spin of protons~\cite{scopel_DAMA}, or magnetic inelastic dark matter \cite{PhysRevD.82.125011} can be ruled out based on the above single model. Inelastic dark matter models are built straightforwardly where a change in the kinematics of the scattering is derived by modifying properties of the dark matter particle. Exploring inelastic scattering scenarios requires experiments using heavy nuclei and sensitivity to high-energy nuclear recoils \cite{PhysRevD.104.103032, BROERMAN2021122212}. In addition, sensitivity to spin-independent and spin-dependent couplings would allow testing many inelastic dark matter scenarios. The bubble chamber technology developed by the PICO collaboration is the most straightforward technology operational satisfying these requirements. Moreover, models of inelastic dark matter explaining the DAMA signal and based on the properties of iodine can be tested with the PICO-60 CF$_{3}$I bubble chamber. This work establishes limits for inelastic dark matter using fluorine and iodine in the PICO bubble chambers. Limits reported in this work were obtained considering contact operators within an effective field theory approach, which is suitable to study any dependence of the interaction on the transfer of momentum, velocity, or spin from either the nuclei or dark matter.

\section{\label{sec:pico60}PICO-60 CF$_{3}$I and C$_{3}$F$_{8}$ bubble chambers}
The PICO collaboration has operated several bubble chambers at the SNOLAB underground facility~\cite{Smith:2012fq} using fluorocarbon fluids as target material. These detectors consist of an inner system composed of a high purity synthetic fused silica jar and stainless steel (SS) bellows inside a SS pressure vessel filled with hydraulic fluid. The inner system is filled with a fluorocarbon material (CF$_{3}$I or C$_{3}$F$_{8}$). The pressure vessel is inside a water tank providing shielding from external background radiation and temperature control. Cameras are used to photograph the chambers for bubble identification, as a trigger, and for position reconstruction. Low-radioactivity piezoelectric transducers are attached to the silica jar registering the acoustic signal produced by the bubble formation. This acoustic signal is used to reject alpha decay backgrounds. Nuclear and electron recoils are calibrated {\it in-situ} using neutron (AmBe and Cf-252) and gamma sources (Co-60 and Ba-133) \cite{PhysRevD.100.082006}. The main advantage of the bubble chambers developed by the PICO collaboration is their sensitivity to nuclear recoils and in parallel, their insensitivity to electron recoil backgrounds. 

The PICO-60 CF$_{3}$I bubble chamber was filled with 36.8 kg of CF$_{3}$I reaching an exposure of 3415 kg-day at varying thermodynamic or "Seitz"~\cite{1958_Seitz} thresholds between 7 and 20 keV and operating between June 2013 and May 2014 \cite{PhysRevD.93.052014}. This detector aimed to search for WIMP-nucleon spin-independent couplings mostly through iodine and WIMP-proton spin-dependent couplings mostly through fluorine.

The PICO-60 C$_{3}$F$_{8}$ bubble chamber was filled with 52.2 kg of C$_{3}$F$_{8}$ reaching exposures of 1167 kg-day at a 3.29-keV Seitz threshold and 1404 kg-day at a 2.45-keV Seitz threshold in two physics runs between November 2016 and January 2017 \cite{PhysRevLett.118.251301} and between April and June 2017, respectively \cite{PhysRevD.100.022001}. The most stringent direct-detection constraints to date on the WIMP-proton spin-dependent cross-section at $2.5 \times 10^{-41}$ cm$^2$ for a 25 GeV/c$^2$ WIMP were established~\cite{PhysRevD.100.022001}.

\begin{figure*}[htpb!]
    \centering
    \includegraphics[width=\linewidth]{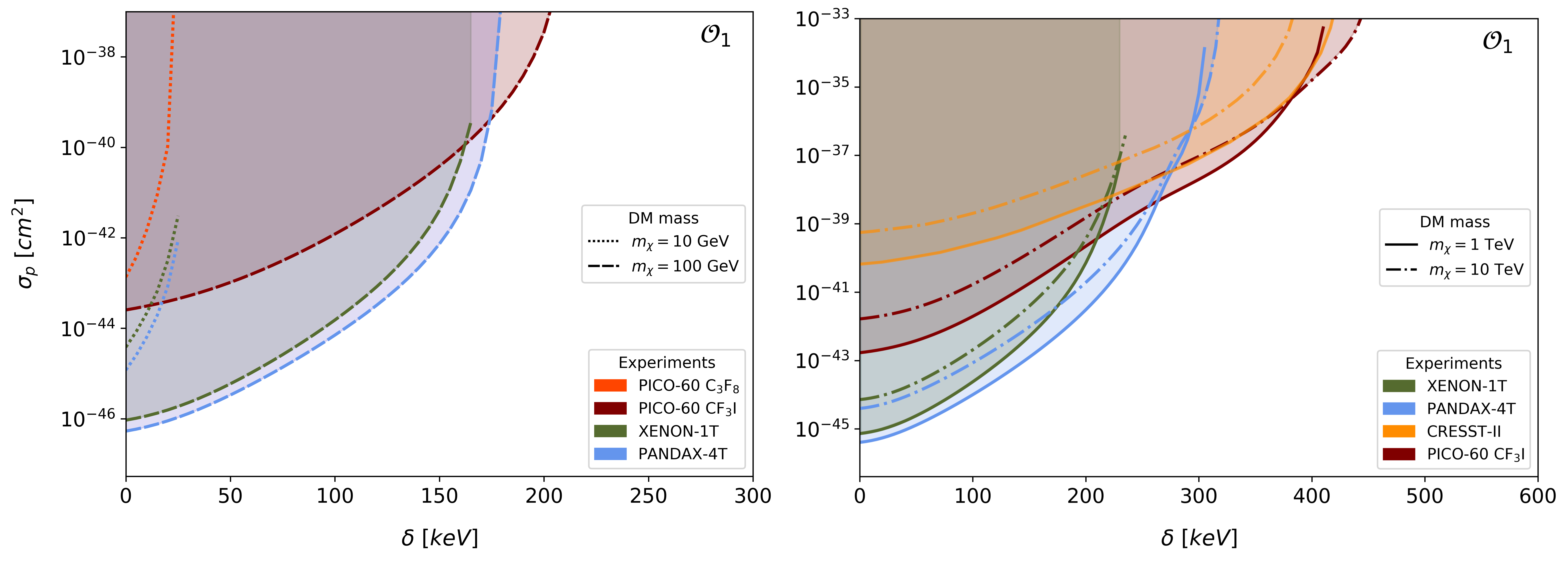}
    \caption{ Upper limits (90\% C. L.) on DM-nucleon scattering cross sections as a function of the mass splitting for the effective operator $\mathcal{O}_1$ and DM masses of 10 GeV/c$^2$ and 100 GeV/c$^2$ (left), and 1 TeV/c$^2$ and 10 TeV/c$^2$ (right), from the analysis of the PICO-60 CF$_{3}$I and C$_{3}$F$_{8}$ experiments. Limits from XENON-1T~\cite{XENON:2018voc}, PANDAX-4T~\cite{PhysRevLett.127.261802}, and CRESST-II~\cite{Kluck_2020} are also shown.}
    \label{fig:O1}
\end{figure*}

The limit calculation method and efficiency curves for both detectors differ since the calibration programs developed by the PICO collaboration were different for each chamber. Namely, a global fit to YBe and AmBe neutron data \cite{Robinson:2015kqr} and pion beam data \cite{PhysRevD.88.021101} was employed to extract the sensitivity of the CF$_{3}$I run. The efficiency curves were then obtained by fitting monotonically increasing, piecewise linear functions \cite{PhysRevD.93.052014}. For the C$_{3}$F$_{8}$ runs, different neutron sources were used, specifically monoenergetic neutrons with energies of 50, 61, and 97 keV produced in $^{51}$V(p,n)$^{51}$Cr reactions, monoenergetic 24 keV neutrons produced by SbBe sources and AmBe neutron data \cite{PhysRevD.100.022001}. Carbon and fluorine efficiency curves for each calibration experiment were obtained similarly as for the CF$_{3}$I run and the efficiency curves for both C$_{3}$F$_{8}$ thresholds were extracted with a fit using the emcee~\cite{Foreman_Mackey_2013} Markov Chain Monte Carlo (MCMC) python code package~\cite{Durnford_2022,PhysRevD.106.122003}.

The exclusion limits shown here use the methods described in \cite{PhysRevD.93.052014} and \cite{PhysRevD.100.022001}. 
The calculations for PICO-60 CF$_{3}$I and C$_{3}$F$_{8}$ followed the standard halo parametrization~\cite{LEWIN199687} assuming a local dark matter density $\rho_D=0.3 $ GeV/c$^2$/cm$^3$ and the same astrophysical parameters for both detectors. The exclusion limits reported are obtained for each of the target fluids employed, both consistent with no dark matter signal. Namely, CF$_{3}$I (variable Seitz thresholds between 7 and 20 keV) and C$_{3}$F$_{8}$ (2.45-keV and 3.29-keV Seitz thresholds). The nuclear recoil energy window extends from the thermodynamic threshold up to 100 keV. The high-energy bound is chosen conservatively due to the absence of acoustic calibration for recoils above $\sim$100 keV.

\begin{figure*}[htpb!]
    \centering
    \includegraphics[width=\linewidth]{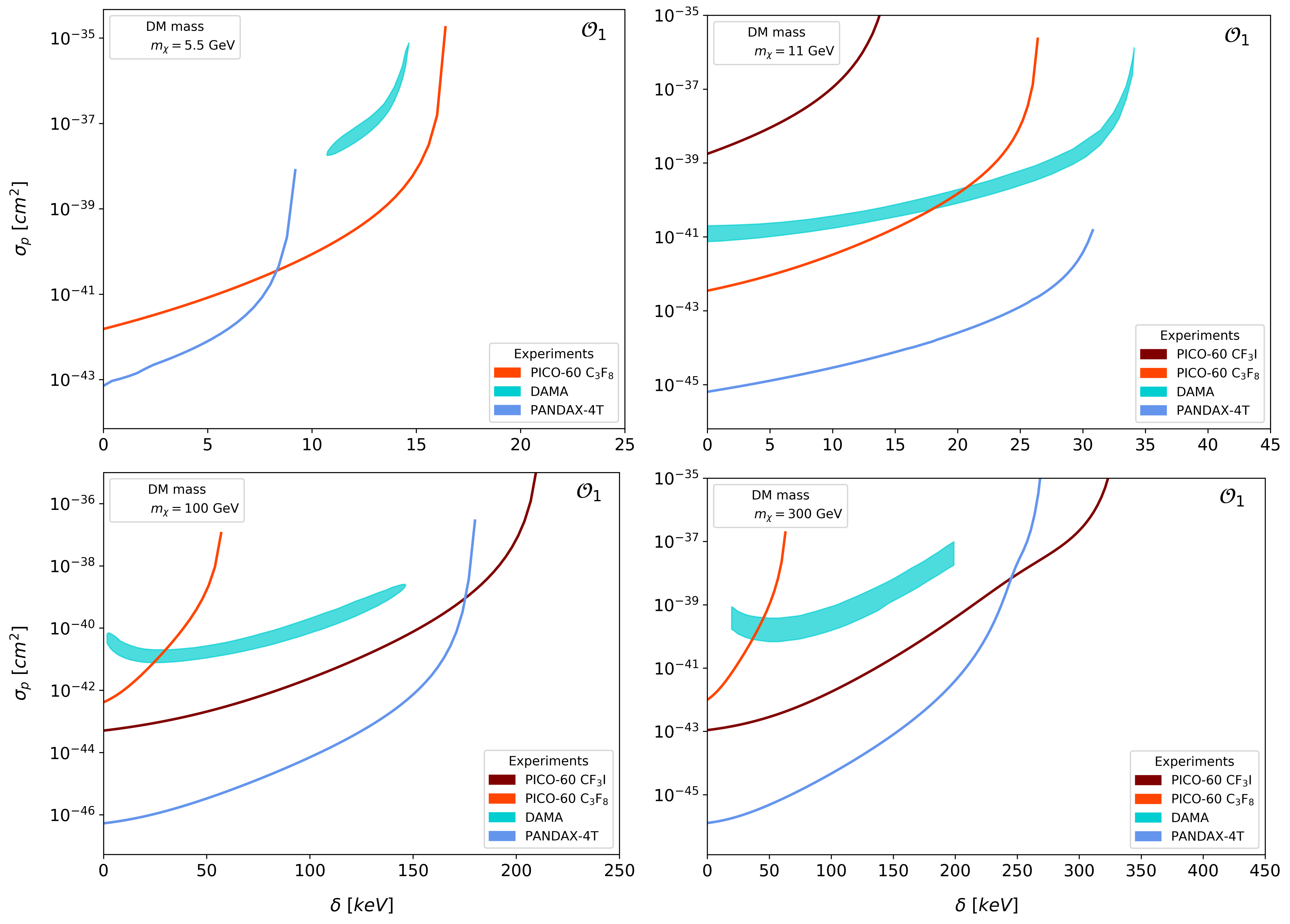}
    \caption{ 99\% C. L. regions allowed by DAMA obtained from~\cite{Schmidt_Hoberg_2009,PhysRevD.64.043502} and PICO-60 CF$_{3}$I and C$_{3}$F$_{8}$ upper limits on DM-proton scattering cross sections as a function of the mass splitting for the effective operator $\mathcal{O}_1$ and DM masses of 5.5 GeV/c$^2$, 11 GeV/c$^2$, 100 GeV/c$^2$, and 300 GeV/c$^2$. Limits from PANDAX-4T are also shown.}
    \label{fig:DAMA_PICO}
\end{figure*}

\section{\label{sec:inelasticDM} Inelastic dark matter}

The observed signal for inelastic scattering is a nuclear recoil constrained by a change in the kinematics of the process with respect to the elastic scattering. The minimum dark matter velocity for the interaction to take place is:  
\begin{equation}
    v_{min}(E_R)=\frac{1}{\sqrt{2M_NE_R}}\left(\frac{M_N}{\mu_{\chi N}}E_R+\delta \right)
\end{equation}
where $\mu_{\chi N}\equiv M_{\chi}M_N/(M_{\chi}+M_N)$ represents the reduced mass of the DM-nucleus system, $E_R$ is the recoil energy, and $\delta$ is the mass splitting between the DM states.
The inelastic scattering is sensitive to higher values of the dark matter velocity. The upper bound to the minimum velocity when compared to elastic scattering reduces the region kinematically accessible. This part of the velocity distribution is more sensitive to the motion of the earth, enhancing the annual modulation~\cite{PhysRevD.64.043502,PhysRevD.72.063509,PhysRevD.79.043513,PhysRevD.94.115026}. The DM-nucleus inelastic interaction could also produce nuclear excitations followed by a de-excitation to the ground state, emitting gamma rays~\cite{ELLIS1988375}. The corresponding response to those nuclear excitations for the isotopes employed in PICO bubble chambers is not considered in this study that rather focuses only in the nuclear recoil response. PICO is insensitive to these interactions, a result of its insensitivity to electron recoils induced by the emitted gamma rays.

A non-relativistic effective field theory (NREFT)~\cite{Fan_2010,fitzpatrick_effective_2013,PhysRevC.89.065501,PhysRevD.92.063515} approach is implemented in this work. This results in quantum mechanical operators depending on exchanged momentum, relative velocity, and nucleon and DM spins. The two operators presented in this work are $\mathcal{O}_1=1_{\chi}1_N$ (with $1_{\chi}$ and $1_N$ as identity operators) and $\mathcal{O}_4=\vec{S}_{\chi}\cdot\vec{S}_N$, where $\vec{S}_{\chi}$ and $\vec{S}_N$ are the DM and nucleus spin, respectively. These two operators are the classical spin-independent and spin-dependent interactions considered by direct detection dark matter experiments. It should be noted that the operator $\mathcal{O}_4$ is particularly significant for inelastic dark matter experiments sensitive to electron recoils, as for odd-mass isotopes low-lying transitions can take place between ground and excited states with different spins~\cite{PhysRevD.61.063503}.


\section{\label{sec:results} Results}

Sensitivity limits as a function of the mass splitting were established for masses of 10 GeV/c$^2$, 100 GeV/c$^2$, 1 TeV/c$^2$, and 10 TeV/c$^2$ for the PICO-60 CF$_{3}$I and C$_{3}$F$_{8}$ bubble chambers. Figure~\ref{fig:O1} shows the upper limits at 90\% C.L. on DM-nucleon inelastic scattering cross-section as function of the mass splitting $\delta$ for operator $\mathcal{O}_1$ (spin-independent coupling). The results reported are compared with limits from XENON-1T~\cite{XENON:2018voc}, PANDAX-4T~\cite{PhysRevLett.127.261802}, and CRESST-II~\cite{Kluck_2020}. The analysis for XENON-1T and PANDAX-4T was performed employing the exposure and background values reported by the collaborations, while results from CRESST-II were obtained from \cite{PhysRevD.94.115026}.

As can be seen in Figure~\ref{fig:DAMA_PICO}, the PICO data excludes the possibility that the DAMA signal is due to an interaction of dark matter through inelastic scattering for mass splittings approximately below 20 keV and above 35 keV~\cite{Schmidt_Hoberg_2009,PhysRevD.64.043502}. For high DM masses ($\sim$ 50 GeV/c$^2$ and above), the PICO-60 CF$_3$I data specifically excludes the interpretation of the DAMA signal as produced by inelastic scattering in iodine. For mass splittings between 20 and 35 keV, PICO excludes all mass ranges considered except for masses near 11 GeV/c$^2$, where only small mass splittings below 20 keV are excluded. In this case, and in general for light DM masses (a few GeV/c$^2$ up to $\sim$ 10 GeV/c$^2$), the contribution from iodine is negligible.
For light DM masses of a few GeV/c$^2$, only lower mass splitting values are experimentally accessible,
$O$(10 keV), and light nuclei, such as carbon and fluorine (PICO), and sodium (DAMA), play an important role in the sensitivity. Although other experiments have also excluded these DM mass ranges and mass splittings, PICO is the only experiment using iodine. In both PICO and DAMA, reaching high mass splittings is possible due to the heavy target (iodine). Bubble chambers have a higher sensitivity to inelastic scattering of dark matter compared to scintillation, phonon and ionization detectors (including DAMA) since they can measure nuclear recoils above the energy threshold and with the potential to extend up to MeV-scale, in contrast to the limited energy recoil window of the other technologies.

\begin{figure}[htpb!]
    \centering
    \includegraphics[scale=0.5]{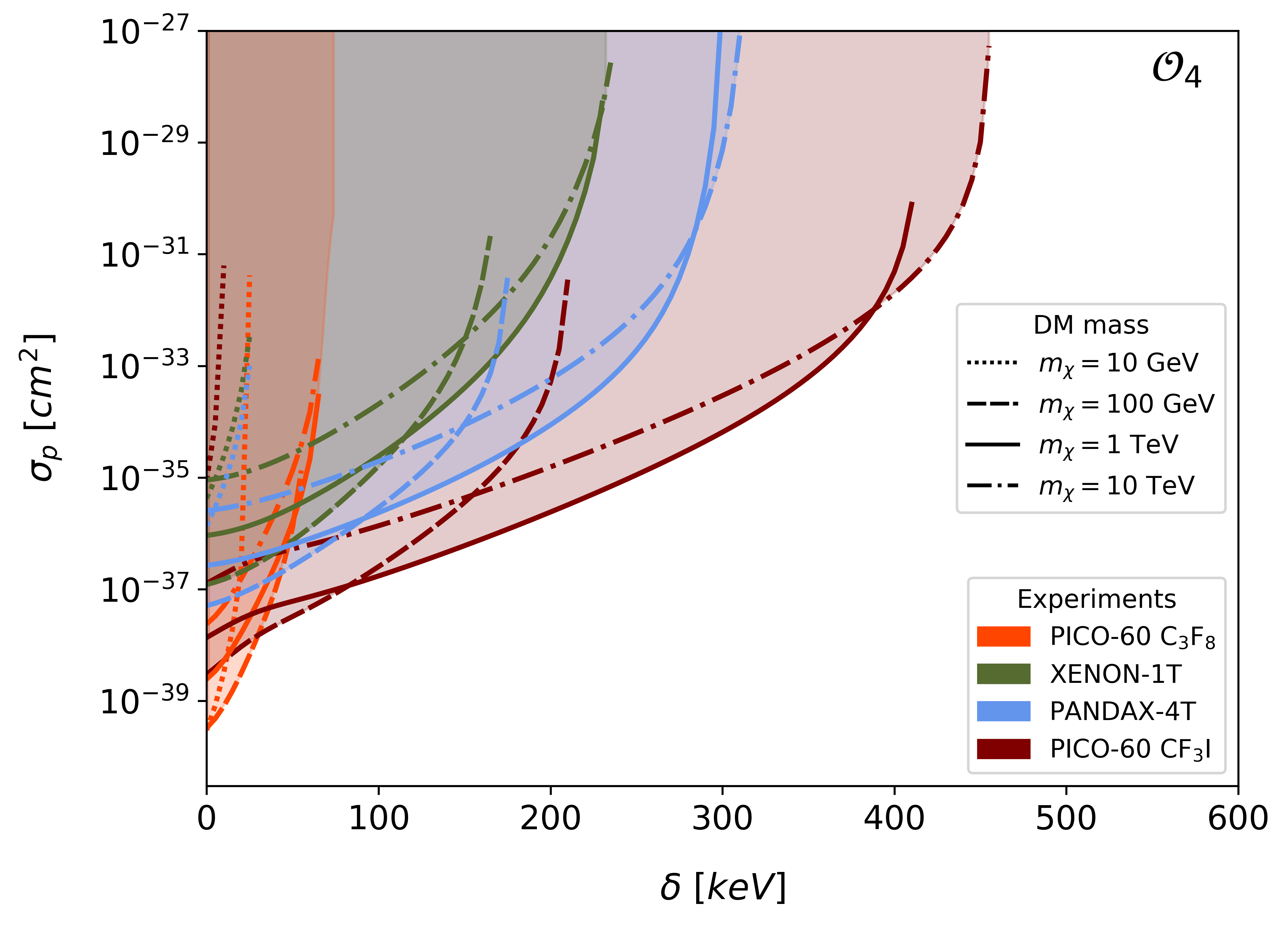}
    \caption{ Upper limits (90\% C. L.) on DM-proton scattering cross sections as a function of the mass splitting for the effective operator $\mathcal{O}_4$ and DM masses of 10 GeV/c$^2$, 100 GeV/c$^2$, 1 TeV/c$^2$, and 10 TeV/c$^2$, from the PICO-60 CF$_{3}$I and C$_{3}$F$_{8}$ experiments. Limits from XENON-1T and PANDAX-4T are also shown.}
    \label{fig:O4}
\end{figure}

Figure~\ref{fig:O4} shows the upper limits at 90\% C.L. on DM-proton inelastic scattering cross-section as function of the mass splitting $\delta$ for operator $\mathcal{O}_4$ (spin-dependent coupling). The results reported are compared with limits from experiments using xenon as the target, namely XENON-1T~\cite{XENON:2018voc} and PANDAX-4T~\cite{PhysRevLett.127.261802}. PICO bubble chambers set leading limits for all possible values of the mass splitting, extending the reach of xenon experiments (PANDAX-4T) above 310 keV and up to approximately 449 keV. PICO-60 C$_{3}$F$_{8}$ dominates for mass splittings below 30 keV, while PICO-60 CF$_{3}$I prevails for values above 30 keV. 

\section{\label{sec:conclusions} Conclusions and discussion}

The results presented in this work establish leading limits on dark matter-nucleon scattering cross-sections for inelastic dark matter interactions in a wide range of mass splittings. The unique sensitivity to high mass splittings results from the combination of the heavy nucleus employed in PICO-60 CF$_3$I and the capability to measure nuclear recoils at all energies above a threshold, contrary to the restricted energy window to measure nuclear recoils by noble, crystal, and semiconductor detectors. The results indicate leading sensitivity to dark matter masses from a few GeV/c$^2$ up to a few TeV/c$^2$ for operator $\mathcal{O}_4$, the classical spin-dependent coupling. These results are relevant since inelastic scattering is useful to distinguish spin-dependent from spin-independent interactions \cite{ELLIS1988375}. In addition, leading limits are reached for operator $\mathcal{O}_1$, the classical spin-independent coupling, for mass splittings between 264 keV and 398 keV (1 TeV/$c^2$ DM mass), as well as between 272 keV and 445 keV (10 TeV/$c^2$ DM mass). This work presents the most sensitive search for inelastic dark matter using fluorine and iodine targets. Based on interpreting the DAMA signal as due to inelastic dark matter interactions, several theoretical scenarios have been mostly excluded by an experiment that, like DAMA, employs iodine. Theoretical scenarios that have been proposed are only allowed for small mass splittings ($\sim$ 10 keV) in a small DM mass window around approximately 10 GeV/c$^2$. Some of these scenarios are proton-philic spin-dependent inelastic dark matter, inelastic scattering predominantly coupling to the spin of protons, and magnetic inelastic dark matter. Iodine has a large magnetic moment that could enhance couplings in models of inelastic dark matter where the DAMA signal could be compatible, but this is mostly excluded by this work. Models~\cite{PhysRevD.79.043513,PhysRevD.64.043502,Schmidt_Hoberg_2009, PhysRevD.80.115008,Kang:2018zld,scopel_DAMA,PhysRevD.82.125011} usually explain the null results from other experiments due to iodine properties, such as magnetic moment or spin of the protons, for example. However the results presented here are based on PICO data that uses iodine as a target as well. PICO bubble chambers continue probing dark matter scenarios with unique sensitivity. 

\section{Acknowledgements}
\begin{acknowledgments}

The PICO collaboration wishes to thank SNOLAB and its staff for support through underground space, logistical, and technical services. SNOLAB operations are supported by the Canada Foundation for Innovation and the Province of Ontario Ministry of Research and Innovation, with underground access provided by Vale at the Creighton mine site. We wish to acknowledge the support of the Natural Sciences and Engineering Research Council of Canada (NSERC) and the Canada Foundation for Innovation (CFI) for funding, and the Arthur B. McDonald Canadian Astroparticle Physics Research Institute. We acknowledge that this work is supported by the National Science Foundation (NSF)
(Grants No. 0919526, No. 1506337, No. 1242637, and No. 1205987), by the U.S. Department of Energy (DOE) Office of Science, Office of High Energy Physics (Grants No. DE-SC0017815 and No. DE-SC-0012161), by the DOE Office of Science Graduate Student Research (SCGSR) award, by the Department of Atomic Energy (DAE), Government of India, under the Centre for AstroParticle Physics II project (CAPP-II) at the Saha Institute of Nuclear Physics (SINP), and Institutional support of Institute Experimental and Applied Physics (IEAP), Czech Technical University in Prague (CTU) (DKRVO). This work is also supported by 
the Project No. CONACYT CB-2017-2018/A1-S-8960, DGAPA UNAM Grants No. PAPIIT IN108020 and IN105923, and Fundaci\'on Marcos Moshinsky. This work is partially supported by the Kavli Institute for Cosmological Physics at the University of Chicago through NSF Grants
No. 1125897 and No. 1806722, and an endowment from the Kavli Foundation and its founder Fred Kavli. We also wish to acknowledge the support from Fermi National Accelerator Laboratory under Contract No. DEAC02-07CH11359, and from Pacific Northwest National Laboratory, which is operated by Battelle for the U.S. Department of Energy under Contract No. DE-AC05-76RL01830. We also thank Compute Canada~\cite{ComputeCanada} and the Centre for Advanced Computing, ACENET, Calcul Qu\'ebec, Compute Ontario, and WestGrid for computational support. The work of M. Bressler is supported by the Department of Energy Office of Science Graduate Instrumentation Research Award (GIRA). 
\end{acknowledgments}
\bibliographystyle{apsrev4-1}
\bibliography{apssamp}

\end{document}